\documentstyle[preprint,aps]{revtex}
\def\ba{\mbox{\boldmath $a$}}
\def\bA{\mbox{\boldmath $A$}}
\def\bx{\mbox{\boldmath $x$}}
\def\bD{\mbox{\boldmath $D$}}
\def\bB{\mbox{\boldmath $B$}}
\def\bk{\mbox{\boldmath $k$}}
\def\bz{\mbox{\boldmath $z$}}

\draft

\begin{document}

\title{Manifestation of Infrared Instabilities in High Energy Processes
in Nonabelian Gauge Theories}

\author{C. Gong$^1$, S. G. Matinyan\footnote{
	On leave from Yerevan Physical Institute, Armenia, 375036
	}$^{1,2}$, B. M\"uller$^1$, and A. Trayanov$^{1,2}$}

\address{$^1$Department of Physics, Duke University, Durham, NC 27708-0305}

\address{$^2$NCSC, Research Triangle Park, NC 27709}

\maketitle

\begin{abstract}
We show that a high frequency standing wave in SU(2) gauge theory
is unstable against decay into long wavelength modes. This provides
a non-perturbative mechanism for energy transfer from initial
high momentum modes to final states with low momentum  excitations.
The Abelian
case does not manifest such instability.
Our analysis is supported by lattice simulations.
\end{abstract}

Recently the intriguing problem of the possibly large rate
for electroweak baryon-number changing processes in high-energy collisions,
originally based on the instanton approach\cite{Ri89}, was considered
from a completely different point of view, based on the classical
analogue to high-energy particle collisions \cite{RT93,Go93}.
The hope was that one could observe the energy transfer from fast (high
frequency) classical wave modes, presumably corresponding to initial
states containing few high-energy particles, to slow (low frequency)
classical wave modes, representing low-energy multi-particle final states.
The studies involved numerical simulations of the (1+1)-dimensional
abelian Higgs model\cite{RT93} and the $\phi^4$-theory in (3+1) space-time
dimensions\cite{Go93}.  In both cases no indication for the existence
of such a mechanism was found.  Instead, the classical high-energy
scattering behavior was found to be fully consistent with expectations
from lowest order perturbation theory.  The nonlinearity of the
investigated theories apparently does not furnish a mechanism for the
formation of final states containing many low-energy particles from the
initial states with few high-energy particles.

In these approaches the main object of study was the nature of the
transfer of energy from high-energy to low-energy modes, which
is thought to be in some sense analogous to the high-energy scattering
situation.  We believe that the observed absence of an efficient energy
transfer mechanism is intimately connected with the integrable nature
of the considered classical systems.
On the other hand, it is well known that nonabelian gauge theories are
nonintegrable in the classical limit
\cite{MST81,CS81,MT92}.
So in these theories there might exist
nonperturbative classical mechanisms for the coupling between fast
and slow modes.
In this paper we want to investigate the existence of such a mechanism
in the
framework of a simple instructive example involving a nonabelian gauge
theory.

Let us model a high-energy particle collision as the interaction between
two counterpropagating Yang-Mills plane waves.
For simplicity let us work in SU(2) gauge theory.
The simplest ansatz satisfying the free Yang-Mills equation is an
abelian standing wave
\begin{equation}
A_i^c(x,t) = \delta_{i3}\delta_{c3} A \cos k_0x \cos \omega_0t
\label{eq01}
\end{equation}
with $k_0=\omega_0$ assumed to be large. (We work in the temporal
gauge $A_0^c=0$.)
We now consider small perturbations $a_i^c({\bx},t)$ around this
solution and study their stability properties.
We will show that there are unstable low-energy modes for arbitrarily
small amplitude $A$ of the driving wave.
The amplitude of these modes grows exponentially with time with a growth
rate proportional to $A$. We will later confirm the presence of the
instability by a numerical calculation and demonstrate that it leads to
rapid total thermalization of the field energy, once the perturbation
has grown to the same strength as the driving wave.

It is easy to see that fluctuations in the same color direction as
the background field ($c=3$) are stable, hence we will consider only
color fluctuations transverse
in color space $a_i =  a_i^1 + i a_i^2$.
The linearized field equations and Gauss' law for the field perturbation are
\begin{equation}
\partial_t^2{\ba} = (\bD\cdot\bD){\ba} - 2i(\bB\times\ba) - \bD(\bD
\cdot\ba)
\label{eq02}
\end{equation}
and
\begin{equation}
\partial_t (\bD\cdot\ba) = 0,
\label{eq03}
\end{equation}
where boldface type indicates three dimensional vectors in position space.
$\bB$ is the magnetic background field generated by the standing
wave, and $\bD=\nabla+i\bA$ is the gauge covariant derivative.
Gauss' law (3) ensures that the physical states are invariant under
time-independent gauge transformations.  To fix  the residual
time-independent gauge, we impose the background field Coulomb gauge
constraint:
\begin{equation}
\bD\cdot\ba = 0.
\label{eq04}
\end{equation}
Now the field equation simplifies, allowing for a decomposition of the
orientations of the polarization vector of the perturbation.
Modes polarized in the direction of the magnetic background field
$\bB$ (the $y$-direction) are stable; hence we only need to consider
modes of the form
\begin{equation}
\ba(\bx,t) = (\hat{\bz} +i\sigma \hat {\bx}) a_\sigma(\bx,t),\qquad
(\sigma=\pm 1)
\label{eq05}
\end{equation}
where $\hat {\bx}$ and $\hat{\bz}$
denote unit vectors in the direction of the
wave vector and the polarization of the background field, respectively.
These modes obey the equation
\begin{equation}
\partial_t^2 a_\sigma - \nabla^2 a_\sigma =
(2i\bA\cdot\nabla - \bA^2 - 2\sigma B) a_\sigma.
\label{eq06}
\end{equation}

We now note that the background field $A_i^c$ oscillates very rapidly
in space and time, whereas we are seeking slowly varying
perturbative modes $a_i^c$. The dynamics of these modes is governed
by the time- and space-averaged interaction with the fast background field.
Since we consider the amplitude of the background wave to be small,
i.e. $A \ll k_0$, we will find a solution perturbatively by calculating
the selfenergy of the fluctuating field to second order in the background
field.  After spacetime averaging, we obtain the following expression for the
selfenergy for a fluctuation with wave vector $|{\bk}|=k \ll k_0$ and
frequency $\omega \ll \omega_0$:
\begin{equation}
\bar\Sigma({\bk},\omega) = -{A^2\over 4} \left[ 1 - (\omega^2 - {\bk}^2)
{\omega^2 + k_{x}^2 \over (\omega^2 - k_{x}^2)^2}
\right]. \label{eq07}
\end{equation}
Note that the expression for $\bar\Sigma$ is independent of the sign
of the circular polarization $\sigma$ of the perturbation (\ref{eq05}).
The dispersion relation for the fluctuation modes is obtained from the
poles of the background field propagator
\begin{equation}
G(\bk,\omega)^{-1} = -\omega^2 + \bk^2 - \bar\Sigma(\bk,\omega) .
\label{eq08}
\end{equation}

We shall check that the perturbative solutions also
satisfy the background Coulomb gauge constraint (\ref{eq04}).
Since the equation of motion (\ref{eq06}) is solved
to second order in $A$, it is consistent that we only require
the solutions to
satisfy Gauss' law to second order. For a particular momentum $\bk$,
we look for a solution of the form
\begin{equation}
\ba(\bx,t) = \sum_{\sigma=\pm 1}
(\hat{\bz} +i\sigma \hat {\bx}) a_{\sigma}(\bx,t).
\label{eq09}
\end{equation}
Inserting it into (\ref{eq04}), disregarding the
fast oscillating terms, we get a relation between
$a_{+}$ and $a_{-}$,
which is represented by their
ratio as a function of momentum $\bk$,
\begin{equation}
\frac{a_{+}}{a_{-}}=
\frac{(ik_{x}-k_{z})(1+I_{1})-i I_{2}}{(ik_{x}+k_{z})(1+I_{1})-iI_{2} },
\end{equation}
where
\begin{equation}
I_{1}=\frac{A^{2}}{4}\left( \frac{1}{\omega^{2}-\bk^{2}}-
	\frac{\omega^{2}+k_{x}^{2}}{(\omega^{2}-k_{x}^{2})^{2}}\right) ,
\qquad
I_{2}=\frac{A^{2}}{4} \frac{k_{x}}{\omega^{2}-k_{x}^{2}}.
\end{equation}
Hence for any $\bk$ there is a solution satisfying Gauss' law.
The most unstable modes are those with
$k_{y} = k_{z}= 0$ for which, keeping only terms up to second order
in $A$, we obtain from (\ref{eq08}) the following dispersion relation:
\begin{equation}
\omega^2 = k_{x}^2 \pm {i\over \sqrt{2}}Ak_{x},
\label{eq10}
\end{equation}
which is complex for any non-zero $k_{x}$ and $A$.
We conclude that the infrared instability exists for
arbitrarily small values of the amplitude $A$ of
the background standing wave.
When $k_{x} \gg A$,
the imaginary part of the frequency, i.e. the exponential
growth rate of a perturbation,
\begin{equation}
{\rm Im} [\omega(k)] =  \frac{A}{2\sqrt{2}},
\label{eq11}
\end{equation}
is independent of the wave vector $k_{x}$.

Unfortunately, the convergence of
the above expansion in increasing powers of
$A^{2}$ is not evident.
An explicit calculation shows that the fourth order contribution
to selfenergy has the same form as the second order one (7) near the pole,
but with a smaller numerical factor.
In order to verify the existence of this instability beyond perturbation
theory, and to find out what happens when the fluctuation begins to absorb
a significant fraction of the driving background wave, we have studied the
evolution of a slightly perturbed low-amplitude standing plane wave in SU(2)
gauge theory on a three-dimensional lattice in the classical limit.
The numerical aspects of such simulations have been described in detail
elsewhere\cite{MT92,BGMT93}.  Here we have initialized the gauge field as
an abelian standing wave with an amplitude corresponding to less than 5\%
of the maximal magnetic energy density on the $16^3$ lattice.  When we
apply a small {\it abelian} perturbation, restricted to the same direction in
color space as the standing wave, the field oscillations remain stable,
as shown by the dashed line in Fig. 1. When we add a general
{\it nonabelian}
perturbation, pointing in a random color direction, the field oscillations
develop a visible instability around time $t=20$, as shown by the solid
curve in Fig. 1.

It is instructive to compare the Fourier spectrum of the magnetic field
energy density
\begin{equation}
E_{\rm m}(\bk) = {1\over 2} \int \bB(\bx)^2 {\rm e}^{i\bk\cdot\bx} d^3x,
\end{equation}
at the end of our simulation ($t=50$) with the energy spectrum
of the initial configuration, as shown in Fig. 2, in both cases.
The initial spectrum
is almost completely concentrated in the mode with $|k|=\pi/a$, where
$a$ is the lattice spacing.
The case with a {\it non-abelian} perturbation is shown in the upper part of
Fig. 2, where the final spectrum is distributed over all modes.
On the other hand, in the case with an {\it abelian} perturbation,
the final distribution remains concentrated in the original mode,
as shown in the lower part of Fig. 2. In order to check whether
the available field energy has been thermalized in the case
with non-abelian perturbation, we have calculated the
probability distribution of the magnetic plaquette energy $P(E)$, divided
by the single plaquette phase space $\sigma(E)$. The
distribution $P(E)/\sigma(E)$ falls exponentially with $E$,
indicating thermalization\cite{BGMT93}.

The quantitative feature of the instability
can be characterized by the associated
largest Lyapunov exponent $\lambda_{0}$.
Numerically we find, in dimensionless form,
\begin{equation}
\lambda_0 a=ka f(\alpha), \qquad \qquad \alpha={gAa\over ka},
\label{eq20}
\end{equation}
where $g$ is the coupling constant. Obviously,
$\lambda_0$ survives in the continuum limit $a\to 0$. The function
$f$ is obtained numerically and shown in
Fig. 3. For small $\alpha$ it is a linear function
$f(\alpha) \approx 0.5 \alpha$. Inserting into the above relation,
we have
\begin{equation}
\lambda_0 \approx {gA \over 2},
\end{equation}
which resembles our result (\ref{eq11}) from second order perturbation
but involves a different coefficient.  We note that the scaling behavior
(\ref{eq20}) is different  from the scaling behavior of a Lyapunov
exponent of a random trajectory \cite{BGMT93}, which scales
with the energy rather than the amplitude of the background field.  The
unstable mode described above is more directly related to the
instability of a uniform chromomagnetic background field\cite{NO78},
and can be traced back to the magnetic moment
interaction of a spin-1 particle, expressed by the form
$2 \sigma B$ in (\ref{eq06}).

Let us make one final remark connected with the high-energy limit of eq.
(\ref{eq06}) for the perturbation $\ba(\bx,t)$. If we consider the behavior
of this equation under rescaling of the longitudinal coordinates
$x$ and $t$ (or that of the light-cone coordinates $x\pm t$) by
\begin{equation}
t \to \lambda t,\quad  x \to \lambda x,\quad y \to y,\quad z \to z,
\end{equation}
one will see that after taking the high-energy limit\cite{VV93}
$\lambda\to 0$ the terms with magnetic field
$B$ in (\ref{eq06}), which caused the
above described instability, will disappear.
One is left with the trivial wave equation
\begin{equation}
(\partial^{2}_{t}-\partial^{2}_{x})
	{\ba(\bx,t)} = 0 .\label{eq15}
\end{equation}
This example shows that the transverse directions
must be handled with care because sometimes they can be the origins of
singularities associated with the gauge sector of nonabelian gauge theories.
In our example the color field does not behave like an abelian
electromagnetic field, but shows the complexity of the
transverse dynamics that cannot be ignored.

In conclusion, we have shown that
a standing wave of high frequency in SU(2) gauge theory
is unstable against decay into long wavelength modes.
This implies that interactions between high energy particles
can result in
a final state with many low energy particles, which might
be relevant in baryon-number changing process.

{\it Acknowledgment:} This work has been supported in part by
the U. S. Department of Energy (Grant DE-FG05-90ER40592) and by
a computing grant from the North Carolina Supercomputing Center.
S. G. M. gratefully acknowledges support from the National
Research Council under the CAST program. B. M. thanks S. Mr\'owczy\'nski
for stimulating discussions.

\vfil\eject

{\noindent \Large \bf Figure Captions}
\bigskip

\begin{itemize}
\item[Fig.1:]
The total electric energy as a function of time.  The initial state is
a standing wave with a small perturbation. The solid line shows
the case when the perturbation is non-abelian, which results in the
destruction of the plane wave beginning at $t \approx 20$. The dotted
line is for an abelian perturbation, in which case the standing wave
is stable.  The computation is performed on a $16^3$ lattice and energy
per plaquette is about 0.16 in lattice units.

\item[Fig.2:]
Time evolution of energy spectra. The upper part is for the case of
a non-abelian perturbation, against which the original standing
wave is unstable, while the lower part is for an abelian
perturbation, where we observe no instability.

\item[Fig.3:]
The scaling function $f(\alpha)$ defined in eq.(\ref{eq20}). The
dotted line is a linear fit with slope 0.5. The results are obtained
by varying the amplitude $A$ with a fixed wave number $ka=0.49$
on a $256\times4^2$ lattice.

\end{itemize}

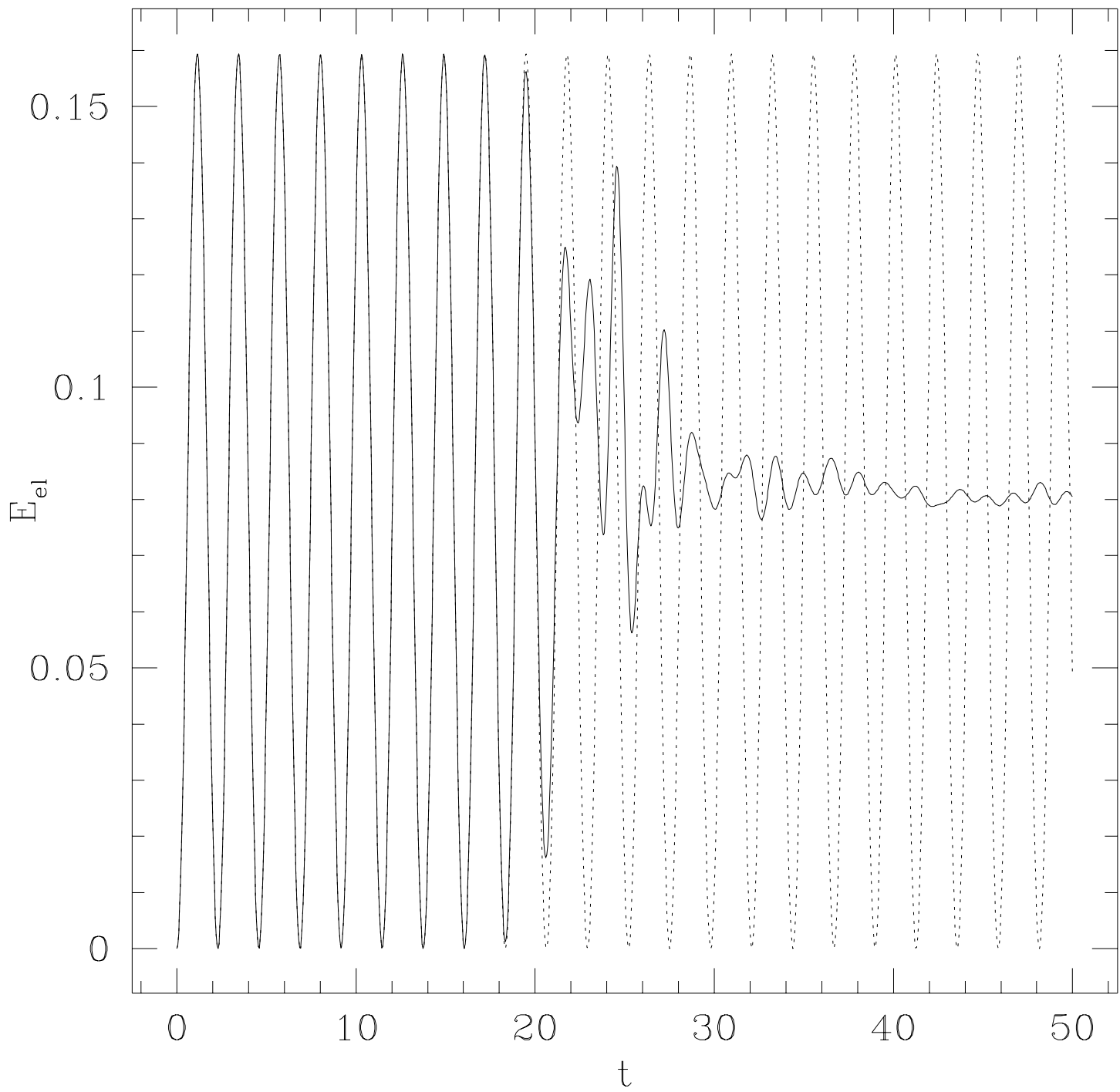
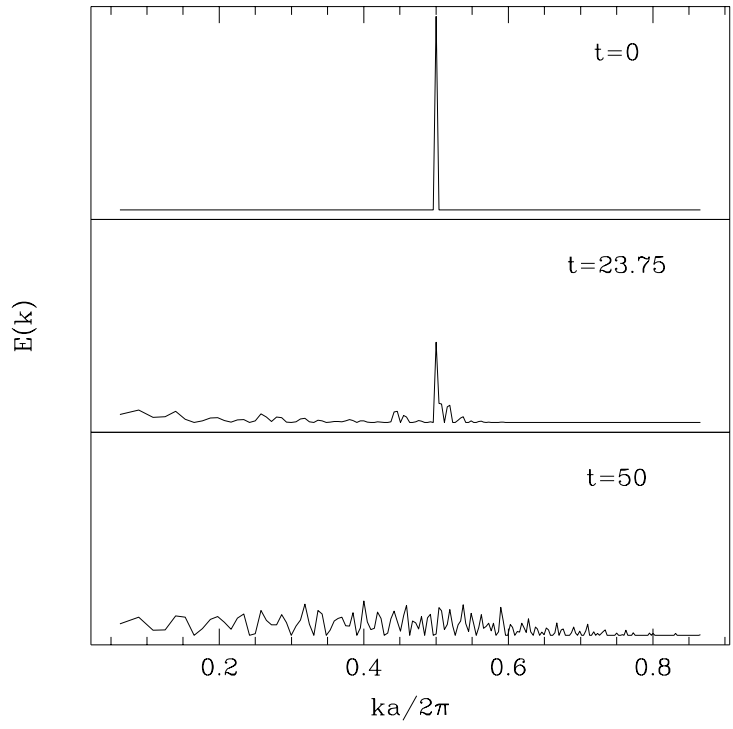
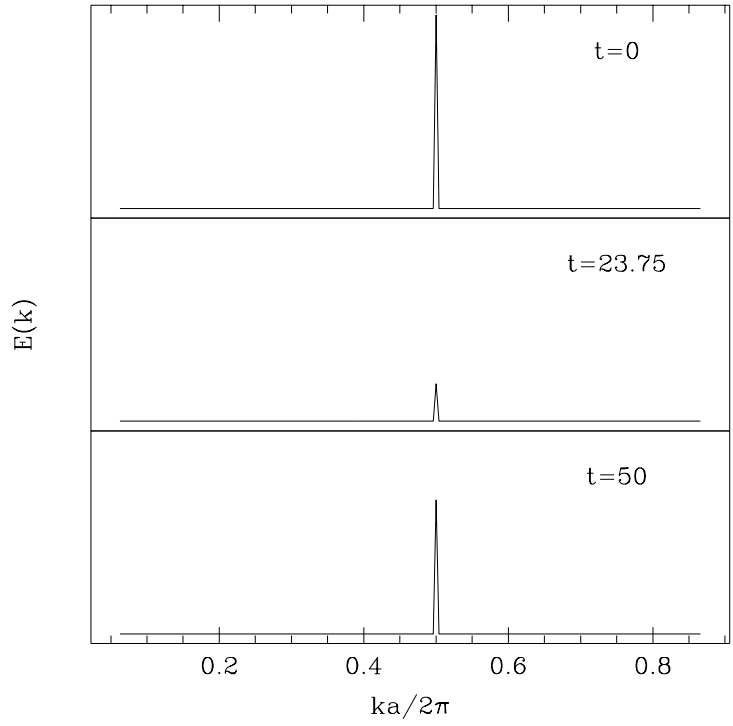
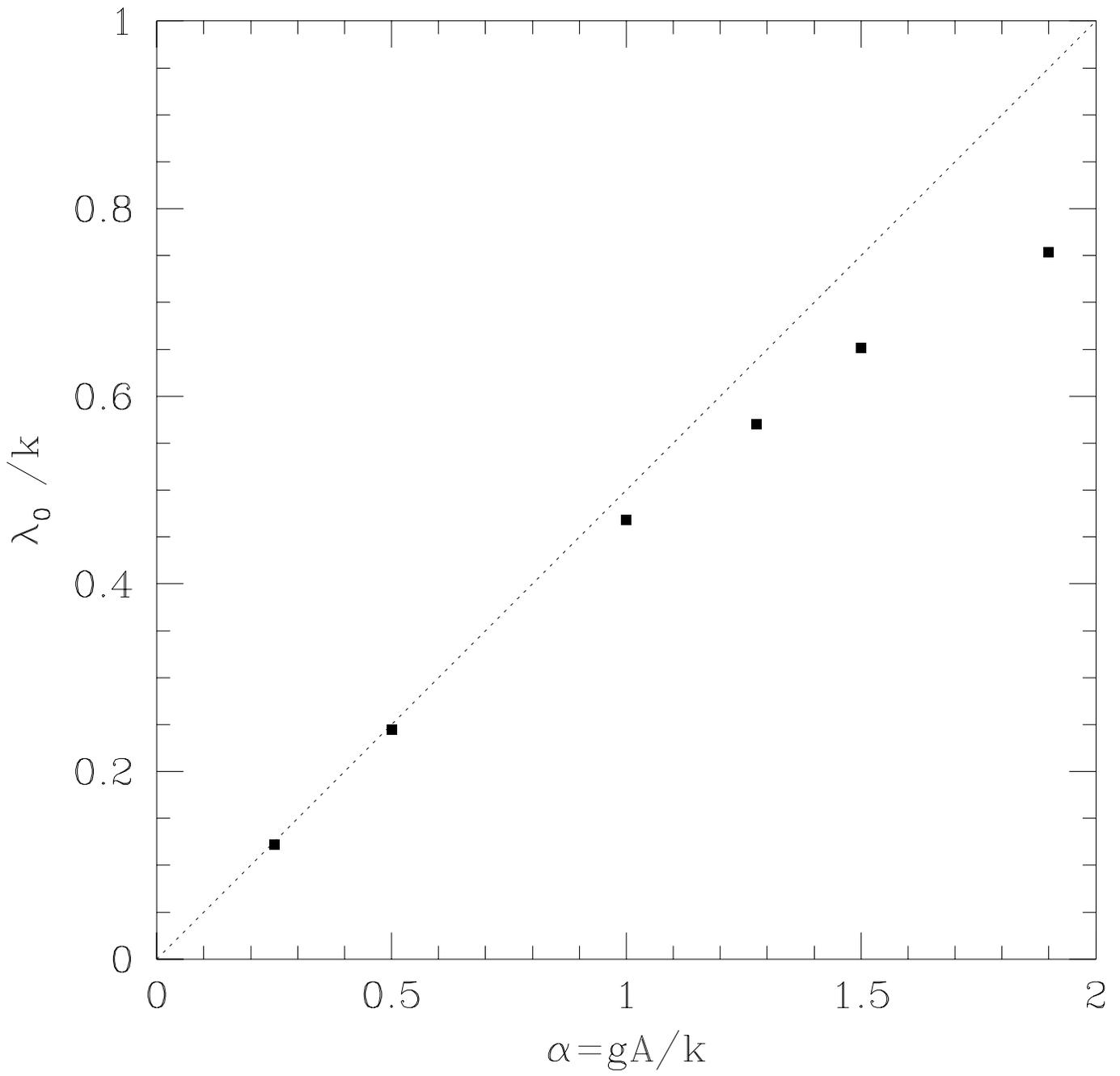
\end{document}